\documentclass[aps,
               pra,
               reprint,
               superscriptaddress,
               nofootinbib,
               twocolumn,
               longbibliography
               ]{revtex4-2}
\usepackage{amsmath,amsthm,amssymb,mathrsfs}
\usepackage{physics}
\usepackage{bbold}
\usepackage[hidelinks]{hyperref}
\usepackage[dvipsnames,table]{xcolor}
  \definecolor{purple}{RGB}{128,0,128}
\usepackage{graphicx}
  \graphicspath{{figs}}
\usepackage{tikz}
  \usetikzlibrary{arrows.meta,calc}
\usepackage[caption=false]{subfig}
\newcommand{\id}{\mathbb{1}}

\begin{document}

\title{The minimal example of quantum network Bell nonlocality}

\author{Erwan Don}
\affiliation{Department of Applied Physics University of Geneva, 1211 Geneva, Switzerland}
\affiliation{Institute for Complex Quantum Systems, Ulm University, 89069 Ulm, Germany}
\affiliation{Center for Integrated Quantum Science and Technology (IQST), Ulm-Stuttgart, Germany}
\author{Jessica Bavaresco}
\affiliation{Sorbonne Universit\'e, CNRS, LIP6, F-75005 Paris, France}
\author{Patryk Lipka-Bartosik}
\affiliation{Center for Theoretical Physics, Polish Academy of Sciences, Warsaw, Poland}
\affiliation{Institute of Theoretical Physics, Jagiellonian University, 30-348 Krak\'ow, Poland}
\author{Nicolas Gisin}
\affiliation{Department of Applied Physics University of Geneva, 1211 Geneva, Switzerland}
\affiliation{Constructor University, Bremen, Germany}
\author{Nicolas Brunner}
\affiliation{Department of Applied Physics University of Geneva, 1211 Geneva, Switzerland}
\author{Alejandro Pozas-Kerstjens}
\affiliation{Department of Applied Physics University of Geneva, 1211 Geneva, Switzerland}

\begin{abstract}
    In recent years, the study of Bell nonlocality has been generalized to quantum networks, where multiple independent sources distribute physical systems to distant parties who perform local measurements.
    In this context, a central open question is to identify the minimal network configuration in which quantum resources produce Bell nonlocal correlations.
    Here we address this question and show that quantum nonlocality is possible in the triangle network where the parties have no input choices and produce only binary-valued outcomes.
    To do so, we start by identifying a family of target distributions and proving their nonlocality.
    Next, we construct an explicit quantum model that reproduces the target distributions to machine precision.
    For this, we develop an efficient method for parameterizing quantum distributions in networks, inspired by the formalism of higher-order quantum operations.
    When considering the number of observed variables and their cardinality, this constitutes the smallest scenario possible that supports quantum nonlocality in networks.
    Moreover, by analyzing the explicit quantum model, we obtain new insights into how nonlocal distributions can be generated in quantum networks.
\end{abstract}

\maketitle

Quantum physics predicts effects that cannot be explained within classical physics.
The paradigmatic example is Bell nonlocality \cite{bell1964,Brunner2014}, where correlations between measurements results in separate systems cannot be accounted for by means of local variables.
Beyond foundations, nonlocality is also the basis for applications, notably within the device-independent framework \cite{Acin2007,ColbeckPhD,Pironio2010}.

\begin{figure*}
	\null\hfill
    \subfloat[]{
        \begin{tikzpicture}
            \draw (2,3.3) rectangle (3,4.3);
            \draw (2.5, 3.8) node {\large $C_c$};
            \draw (0,0) rectangle (1,1);
            \draw (0.5, 0.5) node {\large $A_a$};
            \draw (4,0) rectangle (5,1);
            \draw (4.5, 0.5) node {\large $B_b$};
            \node at (2.5, 0.5) {\LARGE$*$};
            \node at (1.18, 2.22) {\LARGE$*$};
            \node at (3.82, 2.22) {\LARGE$*$};
            \node at (2.5, 0.1) {\large$\rho_\gamma$};
            \node at (0.9, 2.5) {\large$\rho_\beta$};
            \node at (4.2, 2.5) {\large$\rho_\alpha$};
            \draw [-{Latex[length=3mm]}] (2.7, 0.5) -- (4, 0.5);
            \draw [{Latex[length=3mm]}-] (1, 0.5) -- (2.3, 0.5);
            \draw [-{Latex[length=3mm]}] (1.05, 2) -- (0.5, 1);
            \draw [-{Latex[length=3mm]}] (1.3, 2.45) -- (2, 3.7);
            \draw [{Latex[length=3mm]}-] (3, 3.7) -- (3.7, 2.45);
            \draw [-{Latex[length=3mm]}] (3.95, 2) -- (4.5, 1);
            \draw [->] (0.5, 0) -- (0.5, -0.3);
            \node at (0.5, -0.55) {$a\in\{0,1\}$};
            \draw [->] (4.5, 0) -- (4.5, -0.3);
            \node at (4.5, -0.55) {$b\in\{0,1\}$};
            \draw [->] (2.5, 4.3) -- (2.5, 4.6);
            \node at (2.5, 4.85) {$c\in\{0,1\}$};
        \end{tikzpicture}
        \label{fig:triangle:states}
    }
    \hfill
    \subfloat[]{
        \begin{tikzpicture}
            \draw (2,3.3) rectangle (3,4.3);
            \draw (2.5, 3.8) node {\large $C_c$};
            \draw (0,0) rectangle (1,1);
            \draw (0.5, 0.5) node {\large $A_a$};
            \draw (4,0) rectangle (5,1);
            \draw (4.5, 0.5) node {\large $B_b$};
            \node at (2.5, 0.5) {\LARGE$*$};
            \node at (1.18, 2.22) {\LARGE$*$};
            \node at (3.82, 2.22) {\LARGE$*$};
            \node at (2.5, 0.1) {\large$\rho_\gamma$};
            \node at (0.9, 2.5) {\large$\rho_\beta$};
            \node at (4.2, 2.5) {\large$\rho_\alpha$};
            \draw [-{Latex[length=3mm]}] (2.7, 0.5) -- (4, 0.5);
            \draw [{Latex[length=3mm]}-] (1, 0.5) -- (2.3, 0.5);
            \draw [-{Latex[length=3mm]}] (1.05, 2) -- (0.5, 1);
            \draw [-{Latex[length=3mm]}] (1.3, 2.45) -- (2, 3.7);
            \draw [{Latex[length=3mm]}-] (3, 3.7) -- (3.7, 2.45);
            \draw [-{Latex[length=3mm]}] (3.95, 2) -- (4.5, 1);
            \draw [->] (0.5, 0) -- (0.5, -0.3);
            \node at (0.5, -0.55) {$a\in\{0,1\}$};
            \draw [->] (4.5, 0) -- (4.5, -0.3);
            \node at (4.5, -0.55) {$b\in\{0,1\}$};
            \draw [->] (2.5, 4.3) -- (2.5, 4.6);
            \node at (2.5, 4.85) {$c\in\{0,1\}$};
            \filldraw[fill=MidnightBlue, fill opacity=0.3, draw=MidnightBlue, rounded corners]
            (-0.15,-0.1) -- (3.7,-0.1) -- (3.2, 1.25) -- (-0.15, 1.25) -- cycle;
            \filldraw[fill=teal, fill opacity=0.3, draw=teal, rounded corners]
            (3.85, -0.1) -- (5.55, -0.1) -- (3.9, 4.4) -- (2.85, 2.62) -- cycle;
            \filldraw[fill=Mulberry, fill opacity=0.3, draw=Mulberry, rounded corners]
            (0., 1.4) -- (1.94, 1.4) -- (3.72, 4.4) -- (1.78, 4.4) -- cycle;
            \node at (2, -0.4) {\large$R_a$};
            \node at (0.3, 2.7) {\large$T_c$};
            \node at (4.8, 3) {\large$S_b$};
        \end{tikzpicture}
        \label{fig:triangle:testers}
    }
    \hfill\null
	\caption{
        \protect\subref{fig:triangle:states} The binary-outcome triangle network.
        Parties $A$, $B$ and $C$ are pairwise connected by three bipartite sources that distribute quantum states $\rho_\alpha$, $\rho_\beta$ and $\rho_\gamma$.
        Each party performs a measurement on their respective subsystems, resulting in a binary-valued outcome, denoted $a,\,b,\,c\in\{0,1\}$ respectively.
        We characterize the setup by the joint distribution of outcomes, $p(a,b,c)$.
        \protect\subref{fig:triangle:testers} We combine each source with an adjacent measurement in order to characterize the quantum distribution in terms of three independent testers. This enables a more efficient and stable numerical exploration of the set of quantum distributions.
    }
    \label{fig:triangle}
\end{figure*}
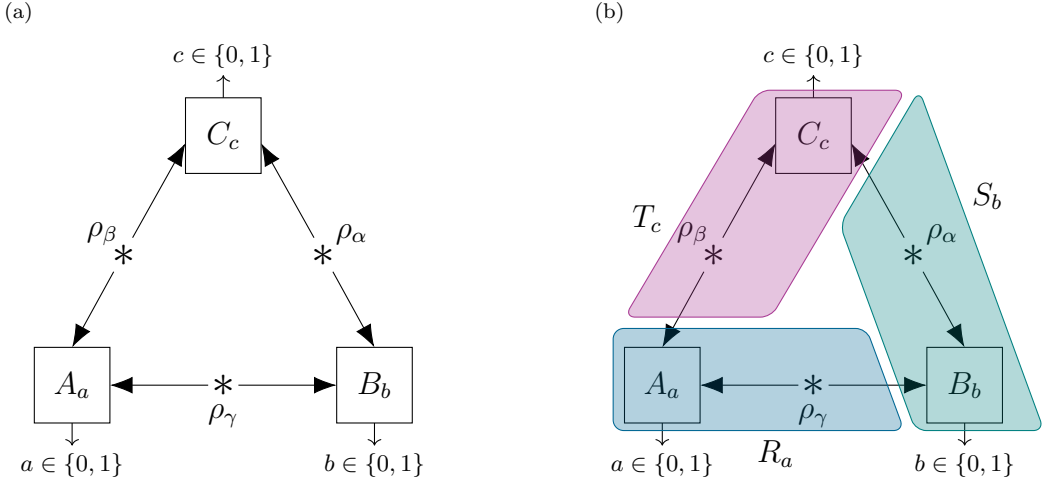

In recent years, growing interest has been devoted to Bell nonlocality in quantum networks (see, e.g.,  \cite{Branciard2010,Branciard2012,fritz2012,Chaves2015,renou2019}, and \cite{networkReview} for a review).
There, multiple independent sources distribute physical systems to distant and non-communicating parties, which can perform joint operations on all the systems they receive (see Fig.~\ref{fig:triangle:states} for a simple example).
From the foundational point of view, networks have unlocked key insights, such as the possibility of generating certified randomness without input choices \cite{sekatski2023,ciudad2025,minati2025}, means to reveal single-photon nonlocality \cite{abiuso2022}, or the importance of complex numbers in quantum theory \cite{renou2021}.

A key question in this area is to identify the minimal scenario that enables quantum network Bell nonlocality.
Minimality can in principle be defined in many different ways, but a natural and commonly considered figure of merit consists of looking for a scenario with the minimal number of observed variables and, in turn, to minimize the cardinality of these variables. 
It has been shown that the minimal scenario needed for demonstrating Bell nonlocality has to involve three observed variables \cite{fritz2012}.
Indeed, any scenario with only two observed variables cannot lead to Bell nonlocality since in this case a classical latent variable can always distribute the values that the visible variables should take at every round. 
This result improves on the standard Bell test, which in its simplest scenario (the Clauser-Horne-Shimony-Holt setup) requires four observed variables (i.e., two input variables and two output variables).
In particular, Bell nonlocality is possible in the so-called triangle network \cite{fritz2012}, sketched in Fig.~\ref{fig:triangle:states}, which features three parties connected by three bipartite sources.
In this case, the three observed variables represent the three outputs of the parties.
Hence, the network features no inputs, in the sense that each party performs a (single) fixed measurement, in contrast to standard Bell tests that require a choice of measurement settings. 

Having identified the minimal number of variables, the next question is to minimize their cardinality.
The first examples of quantum Bell nonlocality in the triangle network involved four-valued outputs \cite{fritz2012,gisin2017,renou2019}.
This could be reduced \cite{Weilenmann2018}, notably to ternary outputs \cite{Boreiri2023}.
However, whether quantum Bell nonlocality in the triangle network could be possible with binary outputs has remained elusive \cite{Gisin2020,bancal2021,Boreiri2023,Pozas2023}, and was even conjectured to be impossible \cite{fraser2018}.

Here we address this question and show that quantum Bell nonlocality is in fact possible in the binary triangle network.
Specifically, we first identify a family of target distributions for which we prove their nonlocality.
Second, we construct an explicit quantum model that reproduces these distributions to machine precision ($\ell_2$ distance of the order of $10^{-16}$).
To do so, we develop an efficient method for constructing quantum distributions in networks, inspired by the formalism of higher-order quantum operations \cite{chiribella2008quantum,chiribella2008transforming,chiribella2009theoretical,taranto2025higher}.
Interestingly, our quantum model requires only one entangled source (the other sources produce classical shared randomness) that distributes two-qubit entanglement, and separable measurements.
Therefore, in addition of representing the minimal example of quantum network Bell nonlocality, our construction requires only basic entanglement resources.

\paragraph*{A family of candidates.---} We consider the ``binary-outcome triangle network'', sketched in Fig.~\ref{fig:triangle:states}.
Three parties, denoted $A$, $B$ and $C$, are connected via three bipartite sources and produce binary outputs, $a,b,c \in \{ 0,1 \}$.
Our focus is on the joint distribution of outputs, $p(a,b,c)$.
This distribution is termed Bell local whenever it admits a decomposition of the form
\begin{equation}
    p(a,b,c)\! = \!\!\!\!
    \sum_{\alpha,\beta,\gamma}\!\!\! q(\alpha)r(\beta)s(\gamma)p_A(a|\beta,\gamma)p_B(b|\alpha,\gamma)p_C(c|\alpha,\beta).
    \label{eq:local}
\end{equation}
where $\alpha$, $\beta$ and $\gamma$ denote classical variables distributed by each source, with their respective probability distributions $q(\alpha)$, $r(\beta)$ and $s(\gamma)$.
Each party provides a binary output according to a local response function given by probability distributions $p_A(a|\beta,\gamma)$, $p_B(b|\alpha,\gamma)$ and $p_C(c|\alpha,\beta)$.
Note that Eq. \eqref{eq:local} naturally generalizes the concept of Bell locality \cite{bell1964} to networks with independent sources \cite{Branciard2010,fritz2012,networkReview}.

Our first step consists in identifying a set of distributions $p(a,b,c)$ that are provably nonlocal, i.e., that do not admit a decomposition of the form \eqref{eq:local}.
We consider the family of so-called noisy $W$ distributions, defined as  
\begin{equation}
    W_v(a,b,c)=\frac{v}{3}\left([001]+[010]+[100]\right) + \frac{1-v}{8},
    \label{eq:W}
\end{equation}
where $[xyz]=\delta_{a,x}\delta_{b,y}\delta_{c,z}$.
Whether the distribution is triangle-local or not depends on the value of the visibility parameter $v$. For $v\leq0.5966$ an explicit triangle-local model has been derived \cite{silva2023}, while for $v>3\left(2-\sqrt{3}\right)\approx 0.8038$ the distribution is provably Bell nonlocal\footnote{This is a consequence of the proof in Ref.~\cite{wolfe2021} that, for higher values of $v$, the corresponding distributions do not admit quantum realizations. Later it was proven that this bound also applied to stronger-than-quantum systems \cite{restivo2022}.}.

We improve on these results by proving that the distributions for $v_1=\frac{622\,070}{1\,000\,000}$, $v_2=\frac{623\,250}{1\,000\,000}$, $v_3=\frac{623\,875}{1\,000\,000}$ and $v_4=\frac{624\,500}{1\,000\,000}$ are nonlocal.
To prove their nonlocality, we use inflation methods \cite{wolfe2019,navascues2017}.
Specifically, we use the implementation in Ref.~\cite{gitton2025} which, crucially, exploits the invariance of $W_v$ under permutations of parties to reduce the computational load\footnote{We use the third level of the corresponding classical inflation hierarchy \cite{navascues2017}, which features three copies of each of the sources, and thus involves solving a linear program for a probability distribution over $2^{27}\sim10^9$ events.}.

It is worth commenting on the fact that the above result is given for specific values of $v$ and not for a continuous interval.
Indeed, intuitively for $v \in [v_1, v_4]$ the distribution should also be nonlocal.
While we believe this is correct (see Ref.~\cite{silva2025} for further support to this intuition), we cannot prove this formally since the sets of local distributions in networks are non-convex.

\paragraph*{Search over quantum realizations.---} 
Now that we have identified candidate distributions that are provably nonlocal, we look for a quantum realization for them.
For this, we develop an efficient numerical optimization method that hinges on concepts from the field of higher-order operations~\cite{ziman08process,chiribella08memory}.
While this method will be discussed in full detail and in a broader context in Ref.~\cite{testers}, here we briefly present the key aspects that are relevant for this work.

For the triangle network without inputs, a distribution admits a quantum model if it can be written as
\begin{equation}\label{eq:quantummodel}
    p(a,b,c)=\text{Tr}\left[\left(\rho_\alpha\otimes\rho_\beta\otimes\rho_\gamma\right)\cdot\left(A_a\otimes B_b\otimes C_c\right)\right]
\end{equation}
where $\rho_\alpha$, $\rho_\beta$, and $\rho_\gamma$ denote the quantum states produced by the sources, and $\{A_a\}$, $\{B_b\}$, and $\{C_c\}$ represent the measurement operators of each party. Note that when computing Eq.~\eqref{eq:quantummodel} one should carefully reorder the subsystems according to Fig.~\ref{fig:triangle:states}. 

Our task consists in showing that distributions in the $W_v$ family, in particular for visibilities $v_1$ to $v_4$, admit a realization of the form \eqref{eq:quantummodel}.
This is a highly non-trivial problem: it is a non-convex optimization problem over six positive-semidefinite matrix variables ($\alpha$, $\beta$, $\gamma$, $A_0$, $B_0$ and $C_0$), and no efficient solving methods are known for this type of problems.
Nevertheless, when five of the variables are fixed, the optimization over the remaining variable is an instance of a semidefinite program (SDP) \cite{sdpReview}, which has a single optimum.
Thus, one can iteratively optimize each of the variables while keeping the remaining ones fixed, performing a \textit{seesaw}-type optimization \cite{werner2001,liang2007}.
Each iteration of such a seesaw algorithm, therefore, requires solving six SDPs.
This process is not guaranteed to converge and, in practice, typically gets stuck in local minima due to the high number of variables to optimize over.

In order to address this, we draw from the field of higher-order quantum operations \cite{chiribella2008quantum,chiribella2008transforming,chiribella2009theoretical,taranto2025higher} the notion of a \textit{quantum tester}~\cite{ziman08process,chiribella08memory}, that describes the objects corresponding to the concatenation of a quantum source and a measurement device.
Formally, given a state $\rho\in\mathcal{L}(\mathcal{H}_1\otimes\mathcal{H}_2)$ and a measurement $\{M_o\}_o$, $M_o\in\mathcal{L}(\mathcal{H}_2\otimes\mathcal{H}_3)$, we define a quantum tester $\{T_o\}_o$, $T_o\in\mathcal{L}(\mathcal{H}_1\otimes\mathcal{H}_3)$ as the set of operators given by
\begin{equation}\label{eq:tester_realization}
    T_o=\text{Tr}_2\left[\left(\rho^{T_2}\otimes\mathbb{1}_3\right)\left(\mathbb{1}_1\otimes M_o^T\right)\right] =: \rho\,*M_o^T,
\end{equation}
where $*$ denotes the link product~\cite{chiribella2008quantum}.
While the characterization of quantum testers as a function of states and measurements is nonlinear, they accept an equivalent description in terms of semidefinite constraints~\cite{ziman08process,chiribella2009theoretical}: A set of operators $\{T_o\}_o$ corresponds to a quantum tester if and only if they satisfy 
\begin{equation}\label{eq:tester}
    T_o \succeq 0 \ \forall \ o, \ \  \text{Tr}\sum_o T_o = d_3, \ \ \sum_o T_o = \left(\text{Tr}_{3} \sum_o T_o\right) \otimes \frac{\id_3}{d_3}.
\end{equation}
Consequently, for any given tester that satisfies Eq.~\eqref{eq:tester}, there exist a quantum state $\rho$ and a measurement $\{M_o\}$ that generate it via Eq.~\eqref{eq:tester_realization}~\cite{ziman08process,chiribella2009theoretical}.

Thus, without loss of generality, it is possible to express any quantum distribution in the triangle network without inputs as
\begin{equation}\label{eq:dist_testers}
    p(a,b,c) = R_a*S_b*T_c,
\end{equation}
where $\{R_a\}$, $\{S_b\}$, and $\{T_c\}$ are independent testers subject to the conditions in Eq.~\eqref{eq:tester} (see Fig.~\ref{fig:triangle:testers}).
This reformulation of the problem restructures the optimization in a crucial way, since the total number of independent matrix variables is reduced.
Thus, each seesaw iteration now requires only three SDP solutions instead of six.
This makes the corresponding seesaw algorithm more stable and more likely to converge to the true optimum.

We implemented this seesaw algorithm minimizing the $\ell_2$ (Euclidean) distance between the target distribution $W_v(a,b,c)$ for different visibilities $v$, and a distribution characterized by quantum testers as in Eq.~\eqref{eq:dist_testers}.
For visibilities up to $v=v_4=0.6245$, we could always find a quantum realization at a small distance to the target.
Next, from the explicit testers returned by the algorithm, we construct an explicit quantum model, that depends on five parameters (see below for details).
Scanning visibilities from $v=0$ to $v=v_4=0.6245$ with a step size of $\frac{1\,249}{1\,998\,000}\sim 6\,{\times}\,10^{-4}$, we find that all corresponding distributions $W_v(a,b,c)$ can be approximated via our model up to machine precision (i.e., $\ell_2$ distance of the order of $10^{-16}$); see Fig.~\ref{fig:alldistance} in Appendix~\ref{app:parameters} and the computational appendix \cite{compapp}, where we showcase results with a finer grid spacing.
Remarkably, we do not observe any notable difference in the $\ell_2$ distance between the regimes where the distribution is triangle-local and when it becomes nonlocal.
From these results, we conclude that Bell nonlocal correlations can be obtained (up to extremely small deviations) via an explicit quantum model.

\paragraph*{The quantum model.---}
\begin{figure}
    \centering
    \begin{tikzpicture}
        \node at (3.33, 5) (c) {};
        \node [draw,shape=rectangle,minimum size=1.33cm,anchor=center] at (c.center) {};
        \node at (0.67, 0.67) (a) {};
        \node (abox) [draw,shape=rectangle,minimum size=1.33cm,anchor=center] at (a.center) {};
        \node at (6, 0.67) (b) {};
        \node (bbox) [draw,shape=rectangle,minimum size=1.33cm,anchor=center] at (b.center) {};
        \node at (1.57, 2.96) (beta) {\includegraphics[scale=.04]{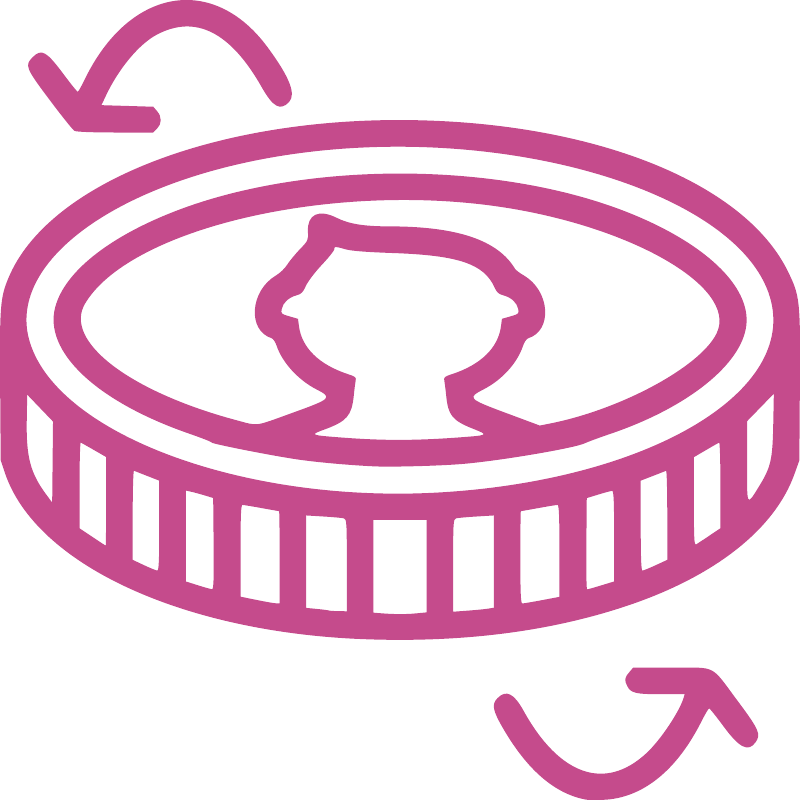}};
        \node at (5.09, 2.96) (alpha) {\includegraphics[scale=.04]{coinMB.pdf}};
        \node at (3.33, 1.15) (flag) {\includegraphics[scale=.04]{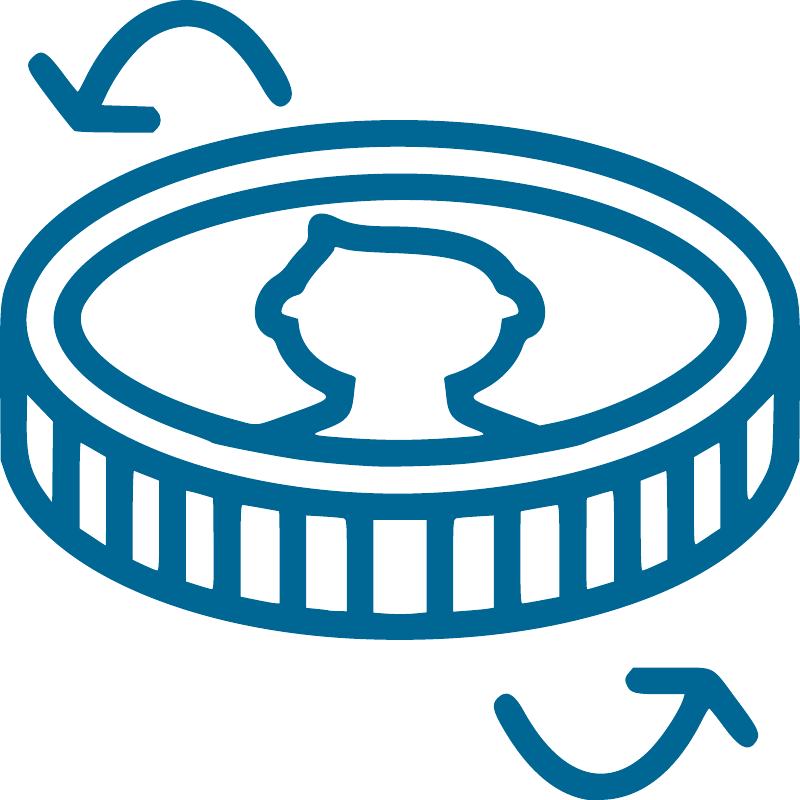}};
        \node at (3.33, 0.6) (gamma) {$\ket{\omega}$};
        \draw [-] (gamma) -- (5.47, 0.6);
        \draw [-] (1.2, 0.6) -- (gamma);
        \draw [color=MidnightBlue] (flag) -- (0.67, 1.15);
        \draw [color=MidnightBlue] (flag) -- (6, 1.15);
        \draw [color=Mulberry,rounded corners] (beta) -- (2.75,5) -- (3.33, 5) -- (3.33, 5.2);
        \draw [color=Mulberry,rounded corners] (beta) -- (0.67,1.33) -- (0.67, 1.15);
        \draw [color=Mulberry,rounded corners] (alpha) -- (3.9,5) -- (3.33, 5) -- (3.33, 5.2);
        \draw [color=Mulberry,rounded corners] (alpha) -- (6,1.33) -- (6, 1.15);
        \draw [->] (0.67, 0) -- (0.67, -0.4);
        \node at (0.67, -0.73) {$a$};
        \draw [->] (6, 0) -- (6, -0.4);
        \node at (6, -0.73) {$b$};
        \draw [->] (3.33,5.12) -- (3.33, 6.1);
        \node at (3.33, 6.4) {$c=(\alpha\oplus1)(\beta\oplus1)$};
        \draw [color=Mulberry,rounded corners,densely dashed] (0.67, 1.15) -- (0.13, 1.15) -- (0.13, 0.13) -- (0.67, 0.13) -- (abox.south);
        \node at (0.8, 0.6) (ma) {\includegraphics[scale=.22]{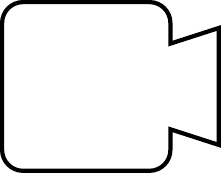}};
        \node at (ma.center) {$M_{\textcolor{Mulberry}{\beta}}$};
        \draw [color=Mulberry,rounded corners,densely dashed] (0.67, 1.15) -- (0.8, 0.92);
        \draw[color=Mulberry,rounded corners,densely dashed] (0.8, 0.28) -- (abox.south);
        \node at (5.87, 0.6) (mb) {\rotatebox{180}{\includegraphics[scale=.22]{meas.pdf}}};
        \node at (mb.center) {$\,\,\,M_{\textcolor{Mulberry}{\alpha}}$};
        \draw[color=Mulberry,rounded corners,densely dashed] (6, 1.15) -- (6.53, 1.15) -- (6.53, 0.13) -- (6, 0.13) -- (bbox.south);
        \draw[color=Mulberry,rounded corners,densely dashed] (6,1.15) -- (5.87, 0.92);
        \draw[color=Mulberry,rounded corners,densely dashed] (5.87, 0.28) -- (bbox.south);
        \fill[black] (0.67, 1.15) circle (2pt);
        \fill[black] (6, 1.15) circle (2pt);
    \end{tikzpicture}
    \caption{
        The quantum model that reproduces the target distributions $W_v$ for visibilities up to $v\leq0.6245$ hence demonstrating nonlocality. The free parameters are the bias of the (pink) coins that connect to party $C$ (the same for both), the bias of the coin that determines whether $A$ and $B$ measure the partially entangled, two-qubit quantum state $\ket{\omega}$, the parameter that characterizes that state, and two more parameters determining the measurements of parties $A$ and $B$ in the X-Z plane.
        In the devices of $A$ and $B$, the coin determines whether they output the coin values received from the other sources or they use them to measure $\ket{\omega}$ and output the results.
    }
    \label{fig:realization}
\end{figure}

Here we present the quantum model in detail.
It turns out to be very simple, and is sketched in Fig.~\ref{fig:realization}.
Two of the sources are in fact purely classical and produce a single bit each, $\alpha,\beta \in \{0,1 \}$, with probabilities $p(\alpha=0)=p(\beta=0)=p_0$.
Upon receiving these two bits, party $C$ outputs $c=(\alpha\oplus 1)(\beta\oplus 1)$, i.e., the outcome $c=1$ happens only when both bits are $0$.
The remaining source distributes an entangled state of the form 
\begin{equation}
    \rho_\gamma =  \left[p_\emptyset \ket{0,0}\bra{0,0} + \left(1-p_\emptyset\right) \ket{1,1}\bra{1,1} \right] \otimes \ket{\omega}\bra{\omega}, 
\end{equation}
which consists of a classical coin with bias $p_\emptyset$, and a two-qubit partially entangled state given by $\ket{\omega}=\cos\omega\ket{00}-\sin\omega\ket{11}$.
The presence of the classical coin is crucial for obtaining nonlocality, as we will discuss below.
Finally, the measurements of parties $A$ and $B$ are given by the observables 
\begin{equation}
    \begin{aligned}
        A=&\,\sigma_z\otimes\ket{0}\bra{0}\otimes\mathbb{1} + \sum_\beta\ket{\beta}\bra{\beta}\otimes \ket{1}\bra{1}\otimes M_\beta,\\
        B=&\,\sigma_z\otimes\ket{0}\bra{0}\otimes\mathbb{1} + \sum_\alpha\ket{\alpha}\bra{\alpha}\otimes \ket{1}\bra{1}\otimes M_\alpha,\\
    \end{aligned}
\end{equation}
where $M_j = \cos{\theta_j} \sigma_z + \sin{\theta_j} \sigma_x $, and the order of the Hilbert spaces corresponds to the source shared with party $C$ (i.e., $\beta$ for party $A$ and $\alpha$ for party $B$) first, the classical coin from $\gamma$ second, and the state $\ket{\omega}$ last.
Intuitively, party $A$'s (respectively, party $B$'s) measurement can be decomposed as follows: (i) check the classical coin from the source $\gamma$, (ii) if its value is $0$, then announce the classical bit received from the source shared with party $C$, i.e., return $a=\beta$ (respectively, $b=\alpha$); otherwise output according to the result of the Pauli observable $M_{\beta}$ (respectively, $M_\alpha$) on the corresponding qubit of $\ket{\omega}$.

A key feature of this construction is the presence of the classical coin in the state $\rho_\gamma$.
To understand why, let us first set $p_\emptyset=1$, in which case parties $A$ and $B$ simply output $a=\beta$ and $b=\alpha$, ignoring the entangled state shared.
Clearly, the resulting distribution is Bell local since all sources produce classical variables.
Second, in the case $p_\emptyset=0$, parties $A$ and $B$ always output according to the Pauli measurements performed on the shared entangled state $\ket{\omega}$.
As we show in Appendix~\ref{app:flagmodel}, the resulting distribution also admits a local model.
Yet, for some intermediate values $0<p_\emptyset<1$, as we saw above, the resulting distribution is nonlocal.
This illustrates a subtle point of nonlocality in networks that is not present in standard Bell scenarios.

Comparing the local models for the distributions with $p_\emptyset=0$ and $p_\emptyset=1$, one can see that they rely on completely different response functions for party $C$.
But since $C$, $\alpha$ and $\beta$ are causally independent from the source $\gamma$, there is no way for them to adapt their internal distributions (in the case of $\alpha$ and $\beta$) and the response function (in the case of party $C$) to the fact that, in a given round, $\gamma$ distributes the state $\ket{0,0,\omega}$ (as in the case when $p_\emptyset=1$) or $\ket{1,1,\omega}$ (as in the case when $p_\emptyset=0$).
Hence these two classical models are incompatible, which explains why the statistical mixture of the two distributions (in the regime $0<p_\emptyset<1$) can exhibit Bell nonlocality.

Additionally, the construction in Fig.~\ref{fig:realization} is reminiscent of the well-known construction by Fritz~\cite{fritz2012}. The latter can be seen as an embedding of a standard CHSH Bell test (with binary inputs and outputs) to the triangle network without inputs.
In Fritz's model, the output cardinality is higher, enabling the effective inputs to be revealed together with the measurement outcomes. 
In our model, in contrast, information about the effective inputs cannot be directly extracted, since the outputs are binary.
Indeed, while Ref.~\cite{fritz2012} establishes a one-to-one correspondence between bipartite nonlocality and network nonlocality, we verify that the corresponding bipartite distribution $p(a,b|\beta,\alpha)$ in our realization is nonlocal for $v>0.495$, while for $v\leq0.5966$ a triangle-local model for the tripartite $W_v$ is known \cite{silva2023}.
Thus, in our model, bipartite nonlocality is a necessary but not sufficient condition for network nonlocality.

\paragraph*{Summary and future directions.---}
We have shown the existence of quantum Bell nonlocality in the binary triangle network, thereby disproving the conjecture of Ref.~\cite{fraser2018}.
We presented a family of quantum models and showed how these can reproduce up to machine precision target distributions that are provably Bell nonlocal. 

We argue that this represents the simplest example of quantum network nonlocality.
Indeed, one can see that trying to further simplify the scenario necessarily prevents the possibility of nonlocality.
Consider first decreasing the cardinality of one output.
Then the corresponding party has a fixed output, and the scenario effectively becomes bipartite, in which case nonlocality is not possible since all correlations can be simulated classically.
Next, consider removing one source.
This results in the ``bilocality'' scenario \cite{Branciard2010,Branciard2012}, which does not support nonlocality when parties receive no inputs\footnote{The proof is straightforward. Consider a distribution $p(a,b,c)$ in the bilocality scenario, hence satisfying $\sum_b p(abc)=p(a)p(c)$. Any such distribution can be simulated classically by having the sources sampling from $p(a)$ and $p(c)$ respectively, and the central party outputting according to $p(b|a,c)$.}.
Therefore, the triangle network with binary outcomes and without inputs, featuring three binary observed variables, is the simplest scenario which supports quantum network Bell nonlocality.
Our result complements recent ones investigating non-classical correlations in more general causal structures, that involve classical communication between the parties \cite{Chaves2017,VanHimbeeck2019}.
In particular it has been shown that, in the bilocality scenario with binary outputs and without inputs, communication allows to witness non-classical constructions despite the fact that all realizable distributions admit local models \cite{lauand2025}.

A key element for obtaining these results is a novel method for efficiently searching for quantum models in networks inspired by the formalism of higher-order quantum operations.
In fact, these techniques have a much broader scope and range of applications which will be discussed in a forthcoming work \cite{testers}. 

We have also discussed in detail the quantum model leading to nonlocal correlations, providing relevant insights as to how nonlocality can arise in networks.
Another relevant aspect is the relative simplicity of the quantum model, which involves only one quantum source, producing simply two-qubit entanglement, while the other two sources are classical and of the smallest possible cardinality.
This construction implies that the distributions that we identify as nonlocal are minimally network nonclassical \cite{ciudad2024}.
An interesting question that remains open is whether the binary-outcome triangle network also allows for more sophisticated forms of network nonlocality, such as full \cite{pozas2022} or genuine \cite{supic2022} network nonlocality. 

A natural question for future work is to derive a fully analytical proof of quantum nonlocality in the binary triangle network, for the distributions we have considered or others.
This could be done by deriving an appropriate Bell-like inequality.
In principle, this is possible via the duality properties of the linear programming techniques which we use throughout this work.
However, the linear programs that we solve here include \textit{linearized polynomial identification constraints} \cite{pozas2022b} (also known as non-certificate-type constraints \cite{Boreiri2023}).
While these allow for significantly improving the characterization of the sets of correlations \cite{alexThesis,Pozas2023,plavala2025}, they prevent the certificate of infeasibility to be interpreted as a Bell-type inequality (see the discussion in Ref.~\cite{pozas2022b}).
This hurdle might be overcome by adapting the techniques described in Ref.~\cite{pozas2022b} to the software solution of Ref.~\cite{gitton2025} that we used.

\paragraph*{Acknowledgments.---}
We thank Tam\'as Kriv\'achy, Sadra Boreiri, Bora Ulu and Antoine Girardin for helpful discussions and comments.
This work is supported by the Swiss National Science Foundation (grant numbers 224561 and 216979, and  NCCR-SwissMAP), the Carl-Zeiss-Stiftung (CZS Center QPhoton), and the Polish National Science Centre through project Sonata 2023/51/D/ST2/02309. Computations were performed in part at the University of Geneva using the Baobab HPC service.

\appendix

\section{Parameterization of the quantum model}\label{app:parameters}
In this section we present in detail the quantum model discussed in the main text and the values of the parameters that reproduce $W_v$ for varying $v$.
The model features five parameters: the probability of $\alpha$ and $\beta$ sending the value $0$ ($p_0$), the probability that $\gamma$ sends the coin in the state $0$ ($p_\emptyset$), the angle that determines the two-qubit state distributed by $\gamma$ ($\omega$), and two angles that characterize the measurements that $A$ and $B$ perform on it ($\theta_0$ and $\theta_1$).

It turns out that some of these parameters can be related to each other and to the visibility $v$ of the target distribution.
By comparing the marginal distribution on the party that receives both classical random variables (i.e., party $C$ in Fig.~\ref{fig:realization}) with the corresponding marginal of $W_v$, it is possible to analytically relate $p_0$ to the visibility of the target distribution $v$, obtaining $$p_0=\sqrt{\frac12-\frac{v}{6}}.$$
Similarly, requiring that the single-body expectation values in the realization coincide for all parties and that the two-body expectation values have the same magnitude and opposite sign than the single-body ones (which happens for all distributions in the $W_v$ family) allows one to express $p_\emptyset$ and $\omega$ in terms of $v$, $\theta_0$ and $\theta_1$ as
\begin{equation*}
    \begin{aligned}
        p_\emptyset&=\frac{v}{3}\frac{\sqrt{6(3-v)}\cos\theta_0+6\cos\theta_1}{(3-v)(\cos\theta_0+\cos\theta_1)},\\
        \cos2\omega&=\!\frac{v\left(6-\sqrt{6(3-v)}\right)}{\left[v\!\left(3+\sqrt{6(3-v)}\right)-9\right]\!\cos\theta_0-9(1\!-\!v)\cos\theta_1}.        
    \end{aligned}
\end{equation*}

\begin{figure}
    \centering
    \subfloat[]{
        \hspace{-5.5pt}\includegraphics[width=0.743\linewidth]{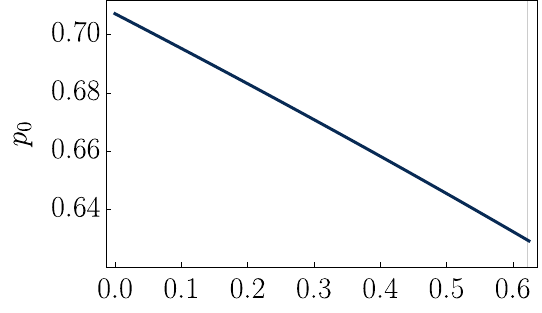}
    }
    \\[-4.7pt]
    \subfloat[]{
        \includegraphics[width=0.72\linewidth]{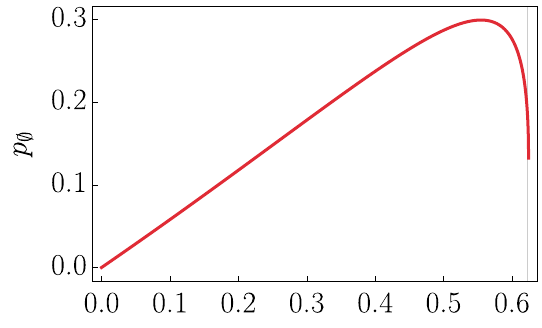}
    }
    \\[-4.7pt]
    \subfloat[]{
        \includegraphics[width=0.72\linewidth]{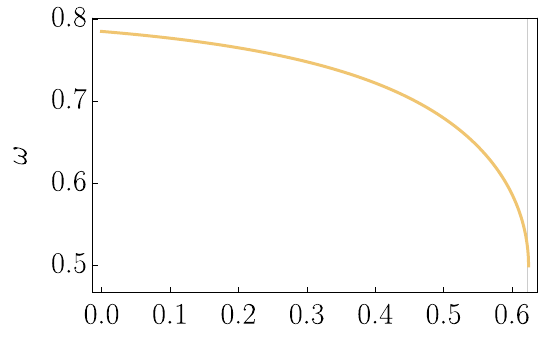}
    }
    \\[-4.7pt]
    \subfloat[]{
        \hspace{-5.5pt}\includegraphics[width=0.743\linewidth]{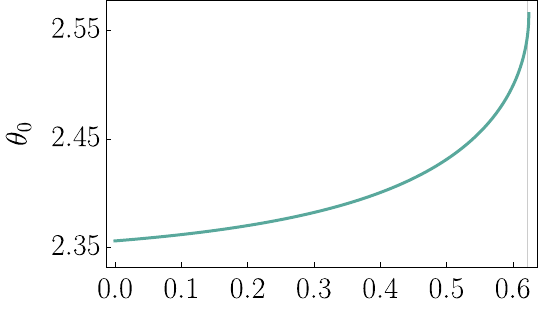}
    }
    \\[-4.7pt]
    \subfloat[]{
        \includegraphics[width=0.72\linewidth]{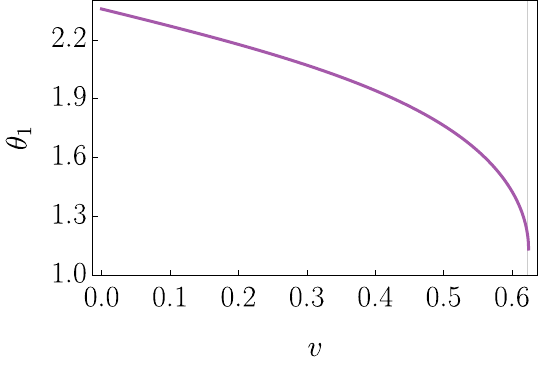}
    }
    \\[-5pt]
    \caption{
        Values of the parameters of the quantum model that reproduces the target distribution $W_v$, as a function of the visibility $v$.
        The parameters $p_0$, $p_\emptyset$ and $\omega$ can be written in terms of the remaining two ($\theta_0$ and $\theta_1$) and the visibility $v$.
        The gray vertical line denotes the value after which the resulting distribution is nonlocal.
    }
    \label{fig:parameters}
\end{figure}

\begin{figure}
    \centering
    \subfloat[]{
        \includegraphics[width=0.95\linewidth]{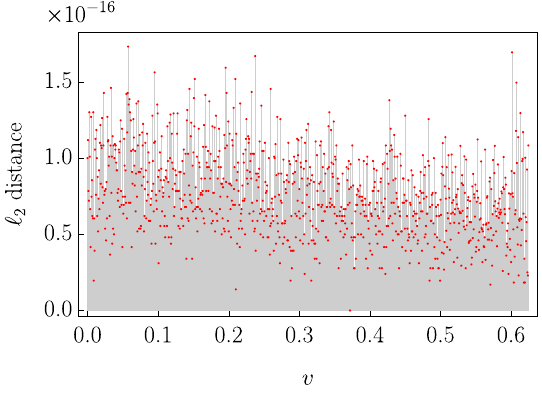}
        \label{fig:alldistance}
    }
    \\
    \subfloat[]{
        \includegraphics[width=0.95\linewidth]{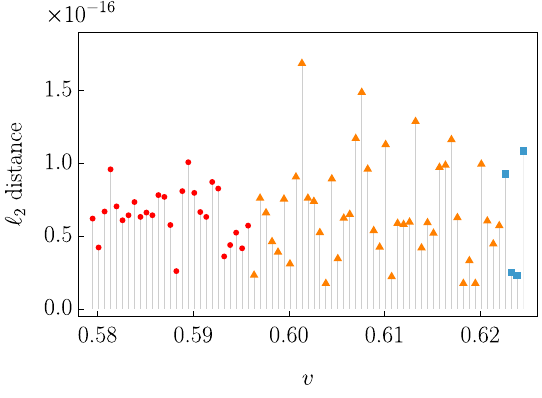}
        \label{fig:zoomdistance}
    }
    \caption{Euclidean distance between the distributions given by the parameters in Fig.~\ref{fig:parameters} and the corresponding $W_v$.
    Panel \protect\subref{fig:alldistance} depicts the distance to models over the interval of visibilities $v\in[0, v_4]$, where $v_4=0.6245$.
    The distributions for which we search for quantum realizations are equally spaced in the interval, with the distance between two consecutive ones being $\frac{1249}{1\,998\,000}$.
    Panel \protect\subref{fig:zoomdistance} is a zoom-in on the region $v\in[0.58, v_4]$.
    There, the red circles correspond to values of $v$ for which the corresponding $W_v$ is known to have a triangle-local model \cite{silva2023}, and the four blue squares correspond to values of $v_1$ to $v_4$ which we prove to be nonlocal. For the intermediate region (orange triangles), we do not know whether the distribution admit a triangle-local model or not.}
    \label{fig:distance}
\end{figure}

In Fig.~\ref{fig:parameters} we plot the values for each of the parameters that reproduce $W_v$ as a function of $v$, for $0 \leq v\leq v_4 = 0.6245$. 
Also, Fig.~\ref{fig:distance} shows the Euclidean distance, measured as
\begin{equation}
    \ell_2(p,q)=\left(\sum_{a,b,c}\left|p(a,b,c)-q(a,b,c)\right|^2\right)^{1/2},    
\end{equation}
between the distribution produced by the model and the corresponding $W_v$.
For these plots, we take uniformly spaced values of $v$, with step size $\frac{1\,249}{1\,998\,000}\sim 6\times 10^{-4}$ between $v=0$ and $v=v_4=0.6245$.
The computational appendix \cite{compapp} also features calculations in the interval $v\in[0.62,0.63]$.
In that case, we use a step size of $10^{-5}$.
There, it is easy to see that, above $v=0.6245$, the distance increases linearly, as expected.

\section{Triangle-local model for $v=0.6245,\,p_\emptyset=0$}\label{app:flagmodel}
To illustrate the importance of the classical coin in the source $\gamma$, we show here that the distribution obtained by omitting the classical coin, i.e., by setting $p_\emptyset=0$ in the model, admits a triangle-local model of the form of Eq.~\eqref{eq:local}.
We set the parameters of the model to those approximating the $W_v$ distribution for $v=v_4=0.6245$, except for $p_\emptyset$, which we set to $0$.
Note that, due to the construction of Fig.~\ref{fig:realization} and the fact that $\ket{\omega}$ is invariant under permutations of the parties, all distributions generated by the strategy satisfy $p(a,b,c)=p(b,a,c)$. 
Thus, the resulting distribution is
\begin{equation}
    \begin{aligned}
        p(0,0,0)=p(0,1,1)=p(1,1,1)=&\,0.054096,\\
        p(0,0,1)=&\,0.233627,\\
        p(0,1,0)=&\,0.258429,\\
        p(1,1,0)=&\,0.033128,
    \end{aligned}
    \label{eq:zeroflag}
\end{equation}
where the remaining probabilities can be obtained from the symmetries.

We use the software developed in Ref.~\cite{silva2023} to perform a numerical search that allows us to infer the following triangle-local model for it.
Let $\alpha$ have cardinality 3 (so it is determined by two parameters, $p_{\alpha=0}\geq0$ and $p_{\alpha=1}\geq0$ that satisfy $p_{\alpha=0}+p_{\alpha=1}\leq1$), and $\beta$ and $\gamma$ have cardinality 2 (so they are determined, respectively, by $0\leq p_{\beta=0}\leq1$ and $0\leq p_{\gamma=0}\leq1$).
Let the response functions be
\begin{equation*}
    \begin{aligned}
        p(a=0|\beta,\gamma)&=\begin{bmatrix}
            x & 1 \\ 1 & \frac{1}{2}
        \end{bmatrix}, \\
        p(b=0|\gamma,\alpha)&=\begin{bmatrix}
        0 & 1 & 1 \\ 0 & 0 & 1    
        \end{bmatrix}, \\
        p(c=0|\alpha,\beta)&=\begin{bmatrix}
            1 & 1 \\ 1 & 0 \\ y & 0
        \end{bmatrix}.
    \end{aligned}
\end{equation*}
For $x=0.131839$, $y=0.294942$, $p_{\alpha=0}=0.082872$, $p_{\alpha=1}=0.738852$, $p_{\beta=0}=0.658567$ and $p_{\gamma=0}=0.571121$, the resulting distribution has an $\ell_2$ distance of $\mathcal{O}\left(10^{-10}\right)$ to that in Eq.~\eqref{eq:zeroflag}.

\bibliography{references.bib}

\end{document}